\documentclass[aps,prfluids,reprint,onecolumn,groupedaddress,12pt]{revtex4-2}
\usepackage{graphicx}
\usepackage{newtxtext}
\usepackage{newtxmath}
\usepackage{hyperref}
\usepackage{comment}
\usepackage[dvipsnames]{xcolor}
\usepackage{textcomp}
\usepackage[normalem]{ulem}
\usepackage{enumerate}
\allowdisplaybreaks[1]

\usepackage{setspace}
\singlespace

\hypersetup{
    colorlinks = true,
    linkcolor  = blue,
    urlcolor   = blue,
    citecolor  = blue,
}

\newcommand{\RomanNumeralCaps}[1]

\definecolor{mygreen}{rgb}{0,0.5,0}
\definecolor{mymagenta}{rgb}{1 0 1}
\definecolor{myblue}{rgb}{0 0 0.75}
\definecolor{myred}{rgb}{0.7 0.11 0.11}

\newcommand{\pd}[3][]{\frac{\partial^{#1} #2}{\partial #3}}
\newcommand{\dd}[3][]{\frac{{\rm d}^{#1} #2}{{\rm d}#3}}
\DeclareMathOperator\erf{erf}

\graphicspath{{Figures/}}

\begin{document}

\title{A theoretical one-dimensional model for variable-density Rayleigh--Taylor turbulence}

\author{Chian Yeh Goh}
\email[]{cgoh@caltech.edu}
\author{Guillaume Blanquart}
\affiliation{Department of Mechanical and Civil Engineering, California Institute of Technology, Pasadena, CA 91125, USA}

\date{\today}

\begin{abstract}
In an early theoretical work published in 1965, \citet{belen1965theory} proposed a turbulent diffusivity model for Rayleigh--Taylor (RT) mixing. We review its derivation and present alternative arguments leading to the same final similarity equation. The original work then introduced an approximation that led to a simplified ordinary differential equation (ODE), which was used primarily to derive the important scaling result, $h\sim (\ln R)gt^2$. Here, we extend the analysis by examining the solutions to both the full similarity ODE and the simplified ODE in detail. It is shown that the full similarity equation captures many now well-known features of non-Boussinesq RT flows, including asymmetric spike and bubble growth and a systematic shift of velocity statistics toward the light-fluid side. Comparisons of the theoretical model with numerical and experimental studies show reasonable agreement in both spatial profiles and growth trends of mixing layer heights. We further show that a global mass correction applied to the simplified solution closely approximates the full solution, highlighting that, to leading order, RT mixing is governed by the competing dynamics between diffusion of $\ln \bar{\rho}$ and mass conservation.
\end{abstract}

\maketitle

\section{Introduction}
\label{sec:intro}

The Rayleigh--Taylor (RT) instability \citep{Rayleigh1882,Taylor1950} arises in a variety of flow applications, including atmospheric science, astrophysics, and inertial confinement fusion \citep{zhou2017rayleigh1}. Between two unstably stratified fluid reservoirs of densities $\rho_H>\rho_L$ under gravity $g$, RT instabilities develop at the interface, and the flow transitions to turbulence. In an infinitely large domain, the flow eventually reaches a late-time state of self-similar growth \citep{youngs1984numerical}, during which the mixing layer height $h(t)$ grows in time as 
\begin{equation}
  h = F(R)gt^2,
  \label{eq:Fgt2}
\end{equation}
where $R = \rho_H/\rho_L$ is the density ratio, $F(R)$ is an unspecified function of $R$, and $t$ is measured relative to a virtual time origin. In the Boussinesq limit, \citet{ristorcelli2004rayleigh} showed that the mixing layer height scales linearly with the Atwood number, $A=(R-1)/(R+1)$, and Eq.~(\ref{eq:Fgt2}) can be written as 
\begin{equation} 
  h \approx \alpha A g t^2,
  \label{eq:Agt2}
\end{equation} 
where $\alpha$ is a dimensionless growth parameter. 

Despite the breakdown of the Boussinesq assumption at finite Atwood number, most variable-density RT studies \citep{Dimonte2000,banerjee2010detailed,youngs2013density,zhou2019time} continue to interpret results using $\alpha$ values defined by Eq.~(\ref{eq:Agt2}). When Eq.~(\ref{eq:Agt2}) is applied separately to the heights of rising bubbles and falling spikes, the bubble growth parameter $\alpha_b$ was found to be largely insensitive to the Atwood number, while the spike growth parameter $\alpha_s$ was found to increase with the Atwood number. As a result, both their sum ($\alpha_T = \alpha_b+\alpha_s$) and their ratio $\alpha_s/\alpha_b$ increase with the Atwood number. Such asymmetries are also reflected in the Favre-averaged velocity and turbulent kinetic energy, which exhibit a systematic shift in their spatial profiles toward the light-fluid side as $A$ increases \citep{Livescu2010,goh2026self}.

Interestingly, a different density scaling for the mixing layer height was proposed much earlier by \citet{belen1965theory}. Using a turbulent diffusivity model, they derived a self-similar solution for the mean density that yields $h \propto (\ln R) g t^2$, which is consistent with Eq.~(\ref{eq:Agt2}) at small $A$. Although this work is commonly credited as an early proponent of $h\sim gt^2$ scaling \citep{abarzhi2010review,youngs2013density,zhou2017rayleigh1,schilling2020progress}, the associated $\ln R$ dependence has, to our knowledge, not been adopted in modern RT studies until it was independently proposed by \citet{goh2026self}. This may be due to the original work being published in Russian and thus less accessible to the broader research community, or due to the rather specific assumptions involved that are difficult to verify without detailed data from high-Atwood-number RT flows. 

The objective of the present work is to revisit the turbulent diffusivity model proposed by \citet{belen1965theory}, examine its assumptions, and use it to address questions that have arisen in the RT literature since the model was first proposed.  Throughout, we clearly distinguish between restatements of the reference work and new contributions introduced here.  Section \ref{sec:math} introduces the turbulent diffusivity model, Sec.~\ref{sec:selfsimilar} details the derivation of the self-similar solution, Sec.~\ref{sec:results} contains further analysis of the model, and Sec.~\ref{sec:discussion} summarizes the physical insights gained.

\section{Mathematical description}
\label{sec:math}
In \citet{belen1965theory}, the mathematical framework was initially formulated for the case of a thin turbulent RT mixing layer superposed on an arbitrary, large-scale background flow. Its presentation involves lengthy derivations, contains a substantial number of typographical errors, and employs a relatively loose notation that does not distinguish between variables associated with the full three-dimensional (3D) equations, the one-dimensional (1D) Reynolds-averaged equations, and the analogous 1D diffusion problem. 

Despite these shortcomings in its presentation, the resulting model has the potential to be highly useful due to its simplicity. Motivated by this, the present section aims to clarify and streamline the original formulation. We focus solely on the canonical RT setup with a quiescent background, thereby significantly reducing the number of variables. In addition, we present simpler, alternative derivations where possible and adopt a clear and consistent notation throughout. Section \ref{sec:math1D} presents the canonical one-dimensional (1D) problem governed by molecular diffusion, while Sec.~\ref{sec:math3D} describes the analogous Reynolds-averaged formulation of the RT problem with turbulent diffusion. Finally, in Sec.~\ref{sec:Dtmodel}, we derive a turbulent diffusivity model that recovers the final form proposed by \citet{belen1965theory}, albeit via slightly different arguments. 

\subsection{Molecular diffusion in one dimension}
\label{sec:math1D}
In one dimension, the temporal variable-density (VD) diffusion problem in $(t,y)$ coordinates is entirely determined by the continuity, scalar transport, and state equations. They are
\begin{gather}
  \pd{\rho}{t} + \pd{\rho v}{y} = 0,
  \label{eq:cont1d}
  \\ 
  \pd{\rho Y}{t} + \pd{\rho v Y}{y} = \pd{}{y}\left(\rho D \pd{Y}{y}\right), 
  \label{eq:scalar1d}
  \\  
  \rho(Y) = \frac{\rho_H \rho_L}{\rho_H- Y\Delta \rho},
  \label{eq:eos1d}
\end{gather}
where $D$ is the molecular diffusivity, $Y$ is the mass fraction, and $\Delta \rho = \rho_H-\rho_L$. These equations can be combined into a single equation for density
\begin{equation}
  \pd{\rho}{t} = \pd{}{y} \left( D \pd{\rho}{y}\right),
  \label{eq:combined1d}
\end{equation}
which admits an analytical solution $\rho = \rho_L + (\Delta \rho/2) [1+\erf(y/\sqrt{4Dt})]$, with the velocity fully determined as  $v = -D (\partial \rho /\partial y)/\rho$.

\subsection{Turbulent diffusion in three dimensions}
\label{sec:math3D}

\citet{belen1965theory} formulated the turbulent RT problem as a diffusive perturbation to a general background flow, before later specializing to the planar configuration with a quiescent background. The resulting 3D equations were subsequently reduced to 1D without any explicit connection between the 1D variables and the underlying 3D flow. Here, we start directly from the canonical RT problem and reduce the 3D equations to 1D using the Reynolds-averaged Navier-Stokes (RANS) framework. 

The Reynolds-averaged continuity, scalar, and state equations for turbulent RT flow (where $\boldsymbol{g} = -g\hat{\boldsymbol{e}}_y$) can be written as 
\begin{gather}
    \pd{\bar{\rho}}{t} + \pd{\bar{\rho} \tilde{v}}{y} = 0,
    \label{eq:cont3d}
    \\
    \pd{\bar{\rho} \widetilde{Y}}{t} + \pd{\bar{\rho} \tilde{v}\tilde{Y}}{y} = \pd{}{y} \left\langle \rho D \pd{Y}{y}\right\rangle -  \pd{}{y} \left( \rho D_t \pd{\tilde{Y}}{y}\right)
    \label{eq:scalar3d}
    \\
    \bar{\rho}(\tilde{Y}) = \frac{\rho_H \rho_L}{\rho_H- \tilde{Y}\Delta \rho},
    \label{eq:eos3d}
\end{gather}
where $\bar{\phi}$ denotes the Reynolds average of $\phi$, $\tilde{\phi} = \overline{\rho \phi}/\bar{\rho}$ denotes the Favre average, and the turbulent scalar flux is expressed in terms of a turbulent diffusivity
\begin{equation}
  D_t = -\frac{\widetilde{vY}-\tilde{v}\tilde{Y}}{\partial \tilde{Y}/\partial y}.
  \label{eq:DtExact}
\end{equation}

Combining Eqs.~(\ref{eq:cont3d})--(\ref{eq:DtExact}), and assuming a sufficiently high P\'{e}clet number (${\rm Pe} \sim h\dot{h}/D$) such that $D_t\gg D$, the equations can be reduced to 
\begin{equation}
  \pd{\bar{\rho}}{t} = \pd{}{y} \left(D_t \pd{\bar{\rho}}{y}\right).
  \label{eq:meandensity3D}
\end{equation}
This is the same mean density equation as (2.21) from \citet{belen1965theory}, and analogous to the 1D diffusion problem described by Eq.~(\ref{eq:combined1d}).

\subsection{Turbulent diffusivity model}
\label{sec:Dtmodel}

The turbulent diffusivity model proposed by \citet{belen1965theory} has a relatively complex derivation, which involves (1) assuming a Prandtl mixing length model, (2) applying Kolmogorov scaling for viscous dissipation, and (3) matching the local entropy contributions from scalar mixing and viscous dissipation, along with additional hydrodynamic and thermodynamic approximations. In the present derivation, we retain the same Prandtl mixing length model, but arrive at the identical turbulent diffusivity model through a more direct line of reasoning. The assumptions are 
\begin{enumerate}[(i)]
    \item The turbulent diffusivity is modeled using Prandtl mixing length theory as 
    \begin{equation}
        D^*_t(t,y) \sim v_t(t,y) h_t(t),
        \label{eq:Dt}
    \end{equation}
    where $h_t$ and $v_t$ are the characteristic length and velocity scales of the turbulence.
    
    \item While $h_t$ is left undefined, the characteristic turbulent velocity $v_t$ is related to $h_t$ by assuming a constant scaling of the global potential energy loss with the kinetic energy of the flow,
    \begin{equation}
      \Delta \rho gh_t^2 = gh_t^2 \int \pd{\bar{\rho}}{y}  {\rm d}y \sim \int \bar{\rho} v_t^2 {\rm d}y.
    \end{equation}
    
    \item It is then assumed that this relationship holds true locally (i.e., at each $y$), yielding 
    \begin{equation}
      v_t \sim \left( \frac{1}{\bar{\rho}}\pd{\bar{\rho}}{y}gh_t^2 \right)^{1/2}.
      \label{eq:vtmodel}
    \end{equation}
\end{enumerate}
Combining Eqs.~(\ref{eq:Dt}), (\ref{eq:vtmodel}), and including a constant of proportionality $K$, the model is
\begin{equation}
  D_t^* = K \left( \frac{1}{\bar{\rho}}\pd{\bar{\rho}}{y}\right)^{1/2} g^{1/2} h_t^2.
  \label{eq:Dtmodel}
\end{equation}

Unlike molecular diffusivity, the turbulent diffusivity $D_t$ is zero in the unmixed reservoirs, and the mean density varies over a compact support $ y \in [-h_s, h_b]$, with boundary conditions
    \begin{gather}
      \bar{\rho}(-h_s) = \rho_L, 
      \qquad
      \bar{\rho}(h_b) = \rho_H, 
      \label{eq:rhoBC1}  
      \\
      \pd{\bar{\rho}}{y}(-h_s) = 0,
      \qquad 
      \pd{\bar{\rho}}{y}(h_b) = 0,
      \label{eq:rhoBC2}
    \end{gather}
where $h_s$ and $h_b$ are not known \emph{a priori} and need to be solved for. In this theoretical model, Eqs.~(\ref{eq:rhoBC1})--(\ref{eq:rhoBC2}) are the definitions of the spike and bubble heights, respectively. The total mixing layer height is defined as $h_T= h_b+h_s$.

\section{Self-similar solution}
\label{sec:selfsimilar}

Using the turbulent diffusivity model proposed in Eq.~(\ref{eq:Dtmodel}), the mean density equation (\ref{eq:meandensity3D}) admits a self-similar solution. The corresponding similarity equation derived by \citet{belen1965theory} is reviewed in Sec.~\ref{sec:sim_eq}. Our primary contribution to Sec.~\ref{sec:selfsimilar} is in Sec.~\ref{sec:scaling}, where we develop a general mapping between similarity and physical-space variables. Finally, the analytical solution obtained by \citet{belen1965theory} for small density ratios is presented in Sec.~\ref{sec:smallR}.

\subsection{Similarity equation}
\label{sec:sim_eq}

This section follows the analysis of \citet{belen1965theory} closely, with some notational changes and added clarification. The turbulent diffusivity model (\ref{eq:Dtmodel}) is used with Eq.~(\ref{eq:meandensity3D}) to yield
\begin{equation}
  \frac{1}{Kg^{1/2}h_t^2}\pd{\bar{\rho}}{t} =  \pd{}{y} \left[\frac{1}{\bar{\rho}^{1/2}} \left(\pd{\bar{\rho}}{y}\right)^{3/2} \right].
  \label{eq:meandensityModel}
\end{equation}
We apply the substitution
\begin{equation}
  \varphi = \frac{\bar{\rho}}{\rho_H},
  \qquad 
  \xi = \frac{y}{h_0},
  \qquad
  {\rm d}\tau = \frac{Kg^{1/2}h_t^2}{h_0^{5/2}} {\rm d}t,
  \label{eq:nondim}
\end{equation}
where $h_0$ is a constant reference length introduced for non-dimensionalization. Equation (\ref{eq:meandensityModel}) can be written in terms of the nondimensional variables as 
\begin{equation}
  \pd{\varphi}{\tau} = \pd{}{\xi} \left[\frac{1}{\varphi^{1/2}} \left( \pd{\varphi}{\xi} \right)^{3/2} \right].
  \label{eq:density_nondim}
\end{equation}
We seek a self-similar solution of the form $\varphi = \varphi(\eta)$ with the similarity coordinate 
\begin{equation}
  \eta = \frac{\xi}{a\tau^s},
  \label{eq:eta}
\end{equation}
where $s$ and $a$ are parameters to be determined. Substituting Eq.~(\ref{eq:eta}) into Eq.~(\ref{eq:density_nondim}), we get 
\begin{equation} 
    -\frac{s\eta}{\tau}\frac{\partial\varphi}{\partial\eta} = \frac{1}{(a\tau^s)^{5/2}}\pd{}{\eta} \left[\left(\pd{\varphi}{\eta}\right)^{3/2}\frac{1}{\varphi^{1/2}}\right].
\end{equation}
For a self-similar solution to exist, $s=2/5$ is necessary to eliminate $\tau$. For notational simplicity of the resulting differential equation, $a=5^{2/5}$ is chosen to yield  
\begin{equation}
  -2\eta\varphi' = \left[\varphi'^{3/2} \frac{1}{\varphi^{1/2}}\right]',
  \label{eq:phiODE}
\end{equation}
with the similarity coordinate
\begin{equation}
  \eta = \frac{\xi}{(5\tau)^{2/5}},
  \label{eq:eta_final}  
\end{equation}
and boundary conditions
\begin{gather}
  \varphi(-\lambda_s) = 1/R, 
  \qquad
  \varphi(\lambda_b) = 1,
  \label{eq:phiBC1}
  \\
  \varphi'(-\lambda_s) = 0,
  \qquad 
  \varphi'(\lambda_b) = 0,
  \label{eq:phiBC2}
\end{gather}
where $\lambda_i$ is the dimensionless mixing layer height in $\eta$ coordinates and $i$ represents bubble, spike, or total height ($\lambda_T = \lambda_b+\lambda_s$). 

Equation (\ref{eq:phiODE}) can be further reduced to a first order ordinary differential equation (ODE) by the substitution $x^2 = \varphi'/\varphi$, resulting in
\begin{equation}
    -2\eta = 3\dd{x}{\eta} + x^3.
    \label{eq:odex2}
\end{equation}
The Neumann boundary conditions in Eq.~(\ref{eq:phiBC2}) become Dirichlet boundary conditions on $x$: 
\begin{equation}
    x(-\lambda_s) =0, 
    \qquad 
    x(\lambda_b) = 0,
    \label{eq:xbc1}
\end{equation}
and the Dirichlet boundary conditions (\ref{eq:phiBC1}) on $\varphi$ become an integral relation on $x^2$:
\begin{equation}
    \int^{\lambda_b}_{-\lambda_s} x^2 {\rm d}\eta = \ln \left(\frac{\rho_H}{\rho_L}\right) = \ln R.
    \label{eq:xbc2}
\end{equation}
If $\lambda_s$ or $\lambda_b$ were known, Eq.~(\ref{eq:odex2}) would be fully determined by specification of either boundary condition in Eq.~(\ref{eq:xbc1}). However, because $\lambda_s$ and $\lambda_b$ are also unknown, solving Eq.~(\ref{eq:odex2}) requires both boundary conditions from Eq.~(\ref{eq:xbc1}) and the additional constraint Eq.~(\ref{eq:xbc2}). Without solving the equations directly, it can be inferred from the form of Eqs.~(\ref{eq:odex2})--(\ref{eq:xbc2}) that $x(\eta; R)$ and all dimensionless heights $\lambda_i(R)$ are functions of $\ln R$.

\subsection{Scaling of physical quantities}
\label{sec:scaling}

Following the derivation of Eqs.~(\ref{eq:odex2})--(\ref{eq:xbc2}), \citet{belen1965theory} proceeded to introduce an approximation that leads to an analytical solution and transformed only that specific result back into physical space. Before reviewing that analysis in Sec.~\ref{sec:smallR}, we contribute a new discussion by considering transformations of the general similarity solution $(\eta,x,\varphi)$ into physical-space variables ($t,y,\bar{\rho},D_t$), allowing for cases where $x$ may be determined analytically or numerically. These general results are then used directly in Sec.~\ref{sec:smallR} and subsequent sections. To facilitate comparisons between similarity variables and physical RT data, equivalent normalizations in both coordinate systems are also established.

The physical coordinate $y$ is related to the similarity variable $\eta$ by Eqs.~(\ref{eq:eta_final}) and (\ref{eq:nondim}). Hence, any physical height $h_i(t,R)$ can be related to its dimensionless form $\lambda_i(R)$ by applying these transformations to yield
\begin{equation}
    h_i(t) = \lambda_i(5\tau)^{2/5} h_0.
    \label{eq:h_working}
\end{equation}
This is clearly valid for the bubble, spike, and total heights, which are explicitly expressed in terms of spatial coordinates. For a self-similar solution, all heights vary proportionally to each other, including the yet-to-be-defined turbulence height $h_t$ (or $\lambda_t$). We apply Eq.~(\ref{eq:h_working}) to $h_t$, substitute it into the equation for ${\rm d\tau}$ (\ref{eq:nondim}), and integrate the resulting expression, leading to
\begin{equation}
    (5\tau)^{1/5} = K \lambda_t^2 \left(\frac{g}{h_0}\right)^{1/2} t.
    \label{eq:tau_t}
\end{equation}
Combining Eqs.~(\ref{eq:h_working}) and (\ref{eq:tau_t}), any physical height of the mixing layer can be written as
\begin{equation}
    h_i = K^2 \lambda_i \lambda_t^4 gt^2.
    \label{eq:h_general}
\end{equation}
This recovers the well-known $h\sim gt^2$ scaling, with density-ratio effects captured by $\lambda_i \lambda_t^4$. Equation (\ref{eq:h_general}) involves two distinct dimensionless heights: the subscript $i$ denotes the specific height being predicted, while the subscript $t$ refers to the turbulence height on which the turbulent diffusivity model (and underlying dynamics) is based. In general, these two heights may take on different definitions and are notated distinctly to reflect this subtlety. In self-similar RT flows, a natural choice for the \emph{normalized} spatial coordinate is 
\begin{equation}
    \frac{y}{h_i} = \frac{\eta}{\lambda_i}.
\end{equation}

The physical density is easily determined from Eq.~(\ref{eq:nondim}) as $\bar{\rho}(t,y) = \rho_H\varphi(\eta)$. Although $\varphi$ is dimensionless, its range varies with the density ratio. To account for density-ratio effects in the normalization, we consider instead the mole fraction $X$, defined as
\begin{equation}
    X = \frac{\bar{\rho}-\rho_L}{\rho_H-\rho_L} = \frac{R\varphi - 1}{R-1},
    \label{eq:molefraction}
\end{equation} 
with values ranging from 0 in the unmixed light fluid to 1 in the unmixed heavy fluid. 

The final quantity of interest is the turbulent diffusivity (\ref{eq:Dtmodel}), which can be written in similarity variables as 
\begin{equation}
    D_t^*  = K \left(\frac{1}{\varphi} \dd{\varphi}{\eta} \pd{\eta}{\xi} \dd{\xi}{y} \right)^{1/2} g^{1/2} h_t^2 
    = \frac{K g^{1/2} h_t^2}{(5\tau)^{1/5}h_0^{1/2}} x.
    \label{eq:Dt_x}
\end{equation}
Combining Eqs.~(\ref{eq:tau_t}), (\ref{eq:h_general}), and (\ref{eq:Dt_x}), the similarity variable $x$ can be interpreted as a dimensionless turbulent diffusivity of the form
\begin{equation} 
  x(\eta) 
  = \frac{D^*_t}{K^4\lambda_t^8g^2t^3} = \frac{2\lambda_i^2}{h_i\dot{h}_i}D^*_t.
  \label{eq:x_Dt}
\end{equation}
Equation (\ref{eq:x_Dt}) shows that $x(\eta)$, though dimensionless, has a density-ratio dependence through the factor $\lambda_i^2$. By explicitly separating the similarity variables from the physical variables, the normalized turbulent diffusivity may therefore be written as
\begin{equation} 
  \frac{D^*_t}{h_i\dot{h}_i} = \frac{x}{2\lambda_i^2}.
  \label{eq:x_norm}
\end{equation}

\subsection{Analytical solution for small density ratios}
\label{sec:smallR}

While Eqs.~(\ref{eq:odex2})--(\ref{eq:xbc2}) can be solved numerically, \citet{belen1965theory} considered an approximation that leads to an analytical solution. For small density ratios, $x^3 = (\varphi'/\varphi)^{3/2}$ is bounded by $(\Delta \rho/\rho_L)^{3/2} = (R-1)^{3/2} \ll 1$. Neglecting the $x^3$ term, Eq.~(\ref{eq:odex2}) can be simplified to 
\begin{equation}
    \dd{\hat{x}}{\eta} = -\frac{2}{3}\eta,
    \label{eq:odex2sim}
\end{equation}
where $\hat{\cdot}$ denotes a solution variable to this simplified ODE. Based on the form of Eq.~(\ref{eq:odex2sim}), $\hat{x}$ must be an even function with boundaries defined by $\hat{\lambda}_s=\hat{\lambda}_b=\hat{\lambda}_T/2$. The solution to Eq.~(\ref{eq:odex2sim}) is easily shown to be
\begin{equation}
    \hat{x} = \frac{1}{3}\left(\frac{\hat{\lambda}_T^2}{4} - \eta^2\right),
    \label{eq:xsim}
\end{equation}
and Eq.~(\ref{eq:xbc2}) simplifies to 
\begin{equation}
    \int^{\hat{\lambda}_T/2}_{-\hat{\lambda}_T/2} \hat{x}^2 {\rm d}\eta 
    = \frac{\hat{\lambda}_T^5}{270} = \ln R,
    \qquad
    {\rm or}
    \qquad
    \hat{\lambda}_T = \left( 270 \ln R \right)^{1/5}.
    \label{eq:eta0sim}
\end{equation}
The density field satisfies $\hat{x}^2 = (\ln \hat{\varphi})'$ and can be integrated analytically as  
\begin{equation}
    \hat{\varphi} 
    = \exp\left\{-\frac{\hat{\lambda}_T^5}{2160}\left[4 - 15 \left(\frac{\eta}{\hat{\lambda}_T}\right) + 40\left(\frac{\eta}{\hat{\lambda}_T}\right)^3 - 48\left( \frac{\eta}{\hat{\lambda}_T}\right)^5\right]\right\}.
    \label{eq:phisim}
\end{equation}
Applying Eq.~(\ref{eq:h_general}) to the total height, assuming $\lambda_t \propto \hat{\lambda}_T$, and substituting Eq.~(\ref{eq:eta0sim}), we get 
\begin{equation}
    \hat{h}_T \propto K^2 \hat{\lambda}_T^5 gt^2 =  270 K^2 (\ln R) gt^2,
    \label{eq:h_smallR}
\end{equation}
which reveals an explicit density-ratio dependence in terms of $\ln R$ instead of the Atwood number, a result consistent with recent work by \citet{goh2026self}. By comparing the integral magnitudes of the neglected term, $|\int (x^3/3) {\rm d}\eta|$, with the RHS of Eq.~(\ref{eq:odex2sim}), $|\int (2\eta/3) {\rm d}\eta|$, \citet{belen1965theory} proposed a range of validity for Eq.~(\ref{eq:h_smallR}) to be $R\lesssim 4$.

\section{Results}
\label{sec:results}

The main result reported by \citet{belen1965theory} is the analytical $\ln R$ scaling (\ref{eq:h_smallR}) derived from the simplified ODE. Besides limited comparisons of $\varphi(\eta)$ profiles between the full and simplified solution, there was no analysis of the height scaling from the full ODE solution or any detailed analysis of the spatial profiles in either case. This section seeks to expand on their work in these two regards. The spatial profiles to both the full and simplified ODEs are examined in Sec.~\ref{sec:profiles} and the scaling of heights are investigated in Sec.~\ref{sec:heights}. In addition, the results from the theoretical model are compared with RT simulations and experiments where possible.  

\subsection{Spatial profiles}
\label{sec:profiles}

\begin{figure}
  \begin{minipage}{0.497\textwidth}
    \includegraphics[width=\textwidth]{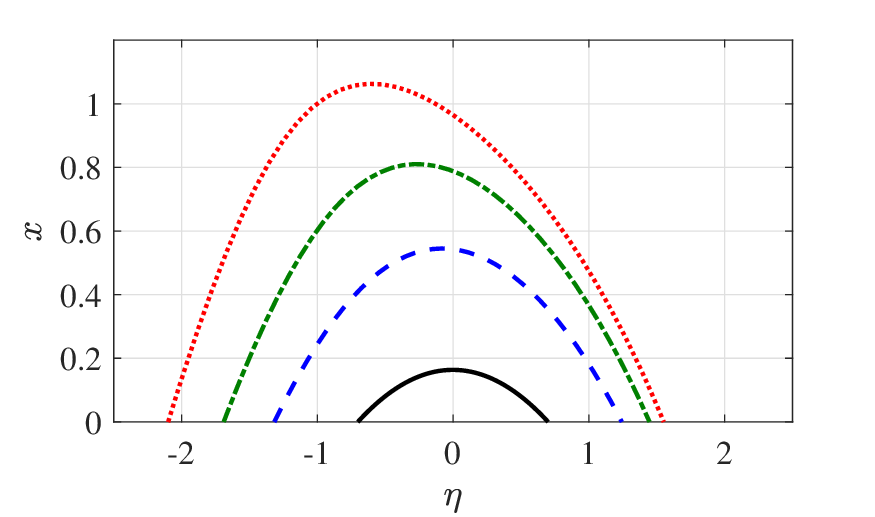}
    {\small (a)}
  \end{minipage}  
  \begin{minipage}{0.497\textwidth}
    \includegraphics[width=\textwidth]{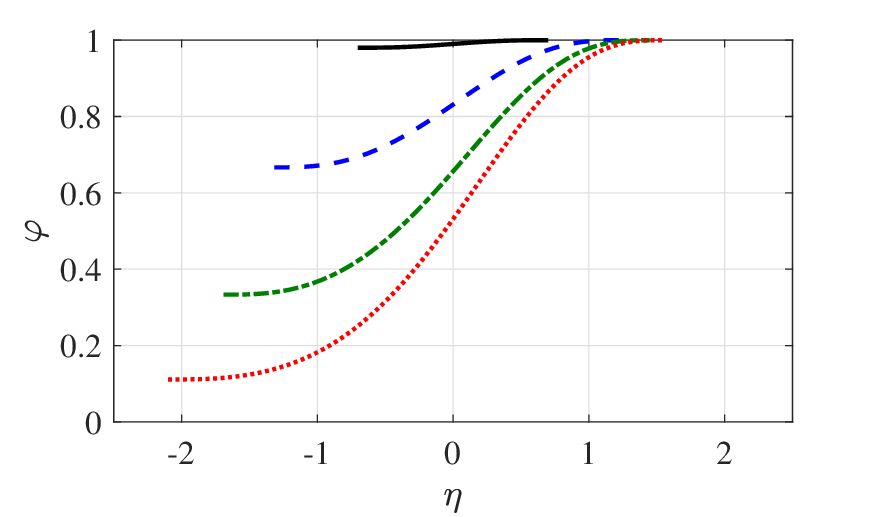}
    {\small (b)}
  \end{minipage}
  \begin{minipage}{0.497\textwidth}
    \includegraphics[width=\textwidth]{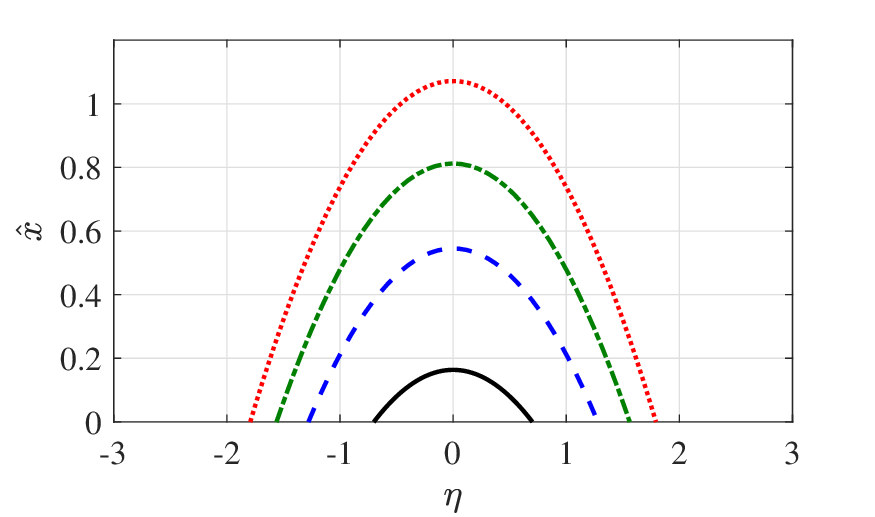}
    {\small (c)}
  \end{minipage}  
  \begin{minipage}{0.497\textwidth}
    \includegraphics[width=\textwidth]{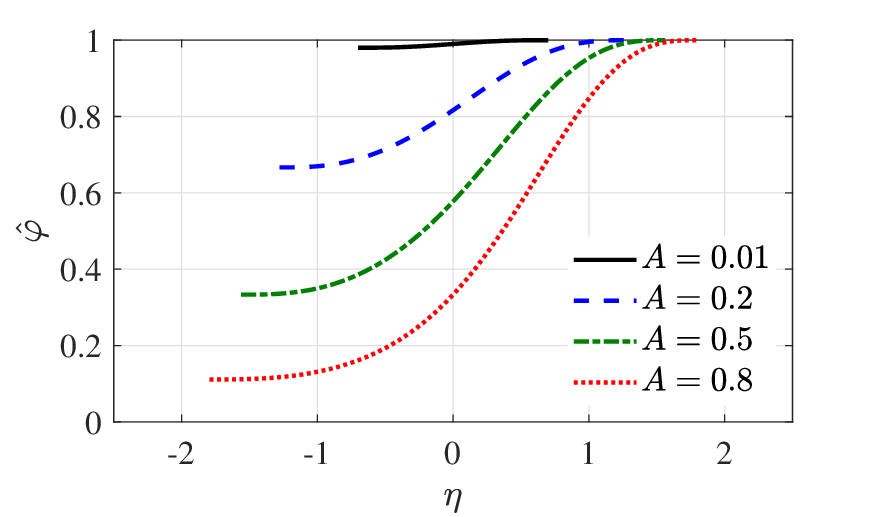}
    {\small (d)}
  \end{minipage}
  \caption{Solutions to full and simplified ODEs. (a,b) Full ODE: dimensionless diffusivity $x = (\varphi'/\varphi)^{1/2}$ and density $\varphi$; (c,d) simplified ODE: dimensionless diffusivity $\hat{x}$ and density $\hat{\varphi}$. Legend: $A=$ 0.01 (black solid), 0.2 (blue dashed), 0.5 (green dash-dotted), and 0.8 (red dotted).
  \label{fig:profiles}}
\end{figure}

The dimensionless diffusivity $x$ and density $\varphi$ for both the full and simplified ODEs are shown in Fig.~\ref{fig:profiles}. While $\eta$, $x$, and $\varphi$ are dimensionless, the systematic growth of the profiles with Atwood number along both axes suggests that the variables are not suitably normalized. For a more effective comparison of their profile shapes, we employ the normalizations proposed in Sec.~\ref{sec:scaling}, where the normalization lengthscale is chosen as $\lambda_T$ (or $\hat{\lambda}_T$). 

\subsubsection{Full ODE}

\begin{figure}
  \begin{minipage}{0.497\textwidth}
    \includegraphics[width=\textwidth]{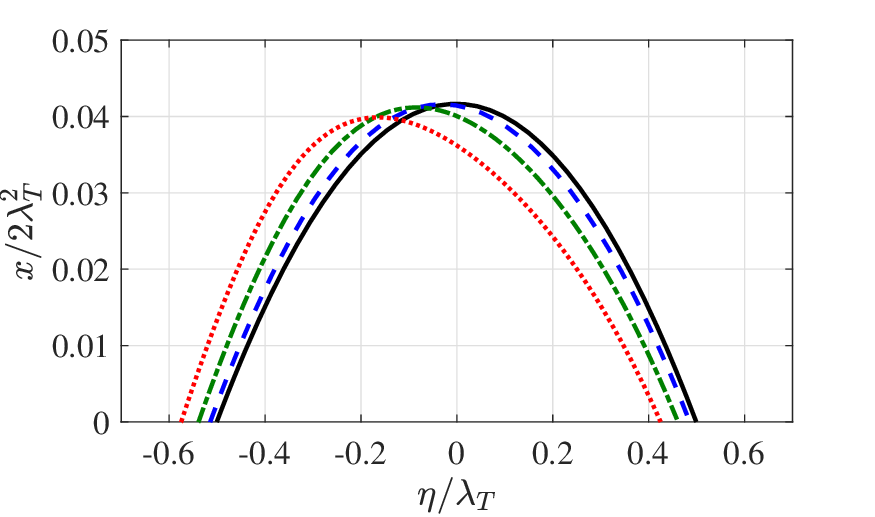}
    {\small (a)}
  \end{minipage}  
  \begin{minipage}{0.497\textwidth}
    \includegraphics[width=\textwidth]{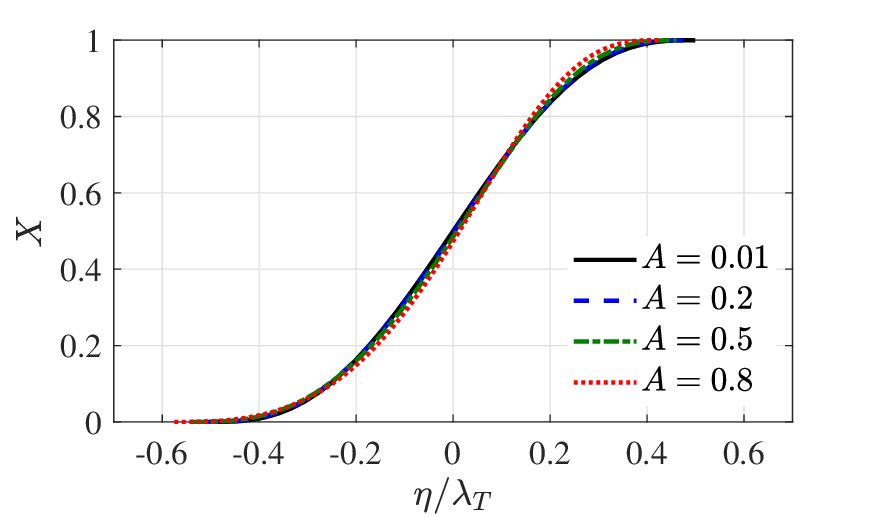}
    {\small (b)}
  \end{minipage}
  \caption{Normalized solution to the full ODE: (a) normalized diffusivity $x/2\lambda_T^2$, and (b) mole fraction $X$. Legend: $A=$ 0.01 (black solid), 0.2 (blue dashed), 0.5 (green dash-dotted), and 0.8 (red dotted).
  \label{fig:profilesNorm}}
\end{figure}

The normalized diffusivity $x/2\lambda_T^2$ from the full ODE (\ref{eq:odex2}) is shown in Fig.~\ref{fig:profilesNorm}(a) and the corresponding mole fraction $X$ in Fig.~\ref{fig:profilesNorm}(b). As the Atwood number increases, three important trends are observed. First, the spike and bubble heights (most clearly identified by the locations where $x=0$) exhibit a pronounced asymmetry with $\lambda_s>\lambda_b$, a now  well-known feature of non-Boussinesq RT flows. Second, noting that $x$ represents not only a turbulent diffusivity but also a turbulent velocity (based on Eq.~(\ref{eq:Dt})), the observed shift of velocity statistics toward the light-fluid side in Fig.~\ref{fig:profilesNorm}(a) is reminiscent of the findings from \citet{Livescu2010}. Third, despite the clear shift in the mixing layer boundaries, the mole fraction collapses fairly well across all Atwood numbers, consistent with the observations reported by \citet{goh2026self}. Remarkably, all three consequences, which have since been observed in the RT literature, are inherently captured by the early theoretical model of \citet{belen1965theory}, but were not mentioned in their original analysis. 

\begin{figure}
  \begin{minipage}{0.497\textwidth}
    \includegraphics[width=\textwidth]{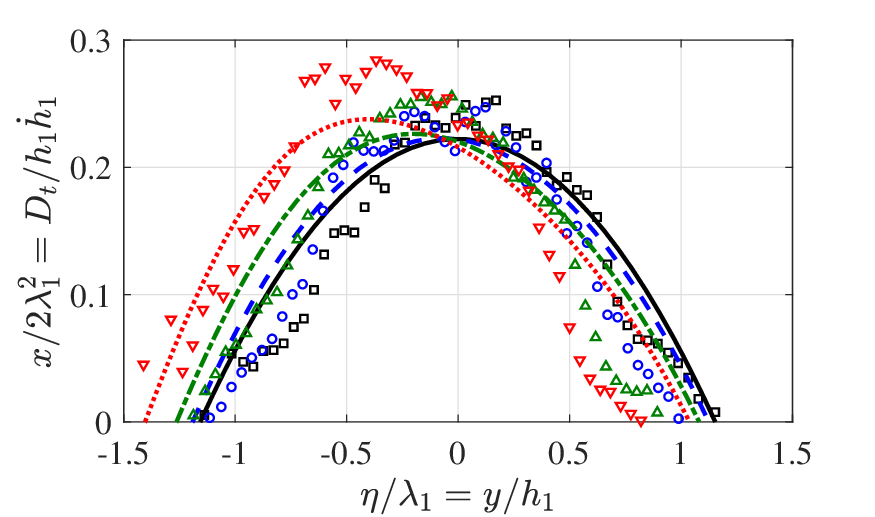}
    {\small (a)}
  \end{minipage}  
  \begin{minipage}{0.497\textwidth}
    \includegraphics[width=\textwidth]{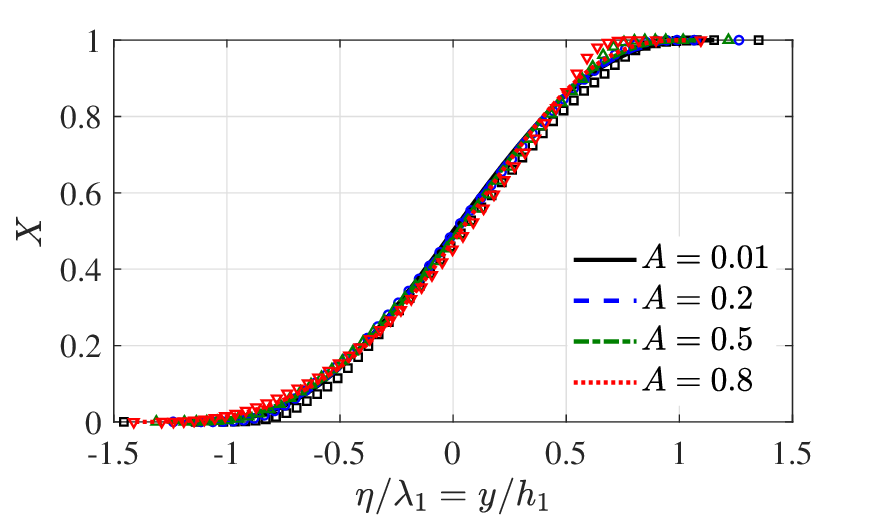}
    {\small (b)}
  \end{minipage}
  \caption{Comparison of the full ODE solution (lines) with DNS results at ${\rm Pe}\approx 1100$ (symbols): (a) normalized diffusivity and (b) mole fraction. Legend: $A=$ 0.01 (black solid, $\square$), 0.2 (blue dashed, {\large $\circ$}), 0.5 (green dash-dotted, $\triangle$), and 0.8 (red dotted, $\triangledown$).
  \label{fig:ModelvSRT}}
\end{figure}

The spatial profiles of the full ODE solution are also compared with DNS data from \citet{goh2026self}. In the theoretical model, the molecular diffusivity is set to $D=0$ and the spike and bubble interfaces are defined unambiguously by Eq.~(\ref{eq:rhoBC1}). In physical RT flows, however, finite molecular diffusivity leads to a continuous variation of the mole fraction across the mixing layer, so that its edges are not uniquely defined. As a result, spike/bubble heights are often defined in terms of mole fraction thresholds, e.g., $h_{s,\theta} = y(X=\theta/100)$ and $h_{b,\theta} = y(X=1-\theta/100)$. To minimize inconsistencies between the infinite-Pe theoretical model and finite-Pe DNS data, an integral height definition $h_1 = 4\int X(1-X) {\rm d}y$ is used for normalization instead. The results from both sources are normalized based on equivalent relations presented in Sec.~\ref{sec:scaling} and shown in Fig.~\ref{fig:ModelvSRT}. In Fig.~\ref{fig:ModelvSRT}(a), the theoretical model for the normalized diffusivity captures the spatial shift observed at higher $A$ and has a reasonable agreement in magnitudes with DNS results after normalization. While some discrepancies remain in the normalized diffusivity, the corresponding mole fraction profiles in Fig.~\ref{fig:ModelvSRT}(b) agree well with DNS results. Deviations are mostly confined to the edges of the mixing layer, where the assumptions underlying the theoretical model are expected to break down.

\subsubsection{Simplified ODE}
\label{sec:profiles_sim}
For the simplified ODE (\ref{eq:odex2sim}), \citet{belen1965theory} reported the analytical $\ln R$ growth for $\hat{h}_T$ in Eq.~(\ref{eq:h_smallR}) without examining the spatial profiles in detail. This section discusses important physical insights that can be gained by analyzing these profiles directly. 

\begin{figure}
  \begin{minipage}{0.497\textwidth}
    \includegraphics[width=\textwidth]{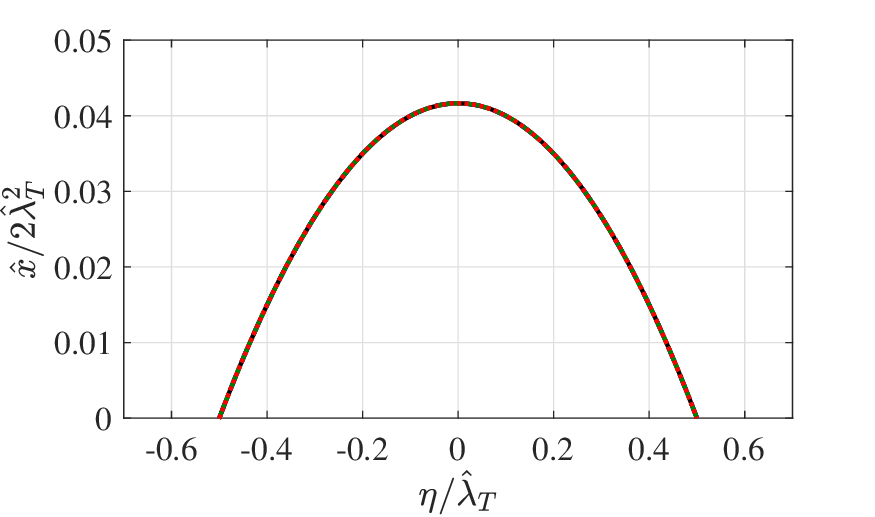}
    {\small (a)}
  \end{minipage}  
  \begin{minipage}{0.497\textwidth}
    \includegraphics[width=\textwidth]{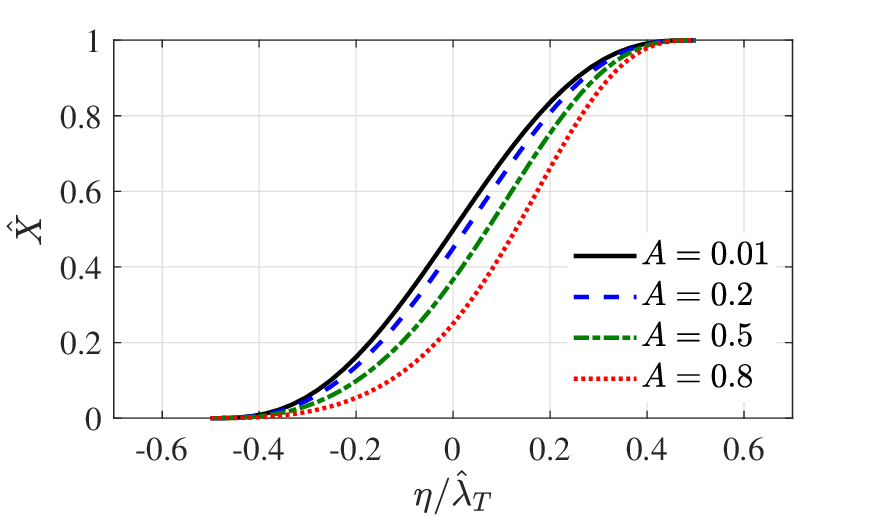}
    {\small (b)}
  \end{minipage}
  \begin{minipage}{0.497\textwidth}
    \includegraphics[width=\textwidth]{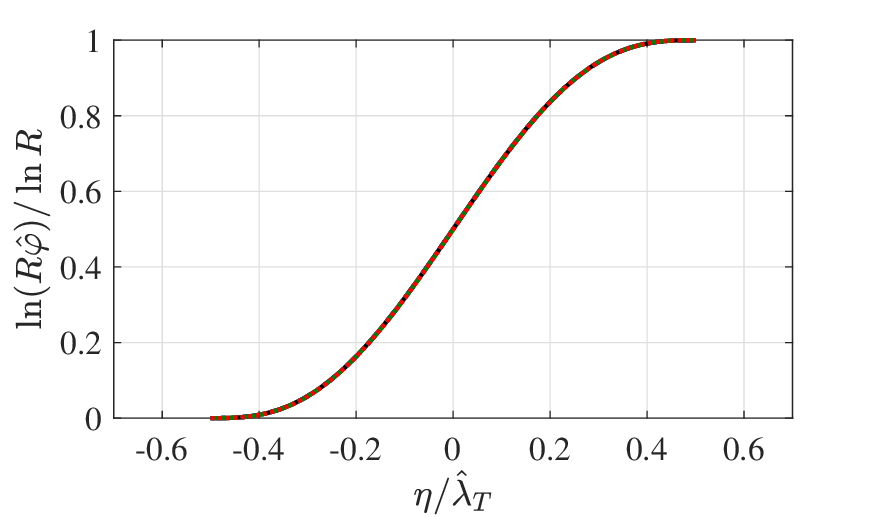}
    {\small (c)}
  \end{minipage}
  \begin{minipage}{0.497\textwidth}
    \includegraphics[width=\textwidth]{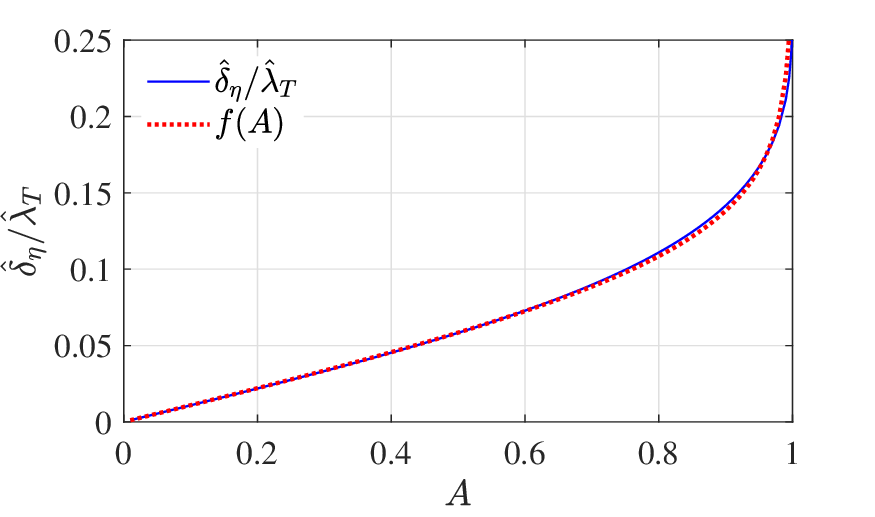}
    {\small (d)}
  \end{minipage} 
  \caption{(a--c) Normalized solution to the simplified ODE: normalized diffusivity $\hat{x}/2\hat{\lambda}_T^2$, mole fraction $\hat{X}$, and normalized $\ln \hat{\varphi}$. Legend: $A=$ 0.01 (black solid), 0.2 (blue dashed), 0.5 (green dash-dotted), and 0.8 (red dotted). (d) Normalized displacement length of the simplified solution, computed from numerically integrating Eq.~(\ref{eq:deltanorm}) (blue solid) and from the analytical expression (\ref{eq:deltanormfit}) (red dotted)
  \label{fig:profilesNorm_sim}}
\end{figure}

The normalized diffusivity $\hat{x}/2\hat{\lambda}_T^2$ and mole fraction $\hat{X}$ are shown in Figs.~\ref{fig:profilesNorm_sim}(a,b), respectively. It is straightforward to show that the analytical form of the normalized diffusivity collapses exactly with Atwood number, as demonstrated in Fig.~\ref{fig:profilesNorm_sim}(a). While $\hat{x}$ is an even function, $\hat{X}(\eta)$ exhibits significant asymmetry within the nominally symmetric interval $[-\hat{\lambda}_T/2,\hat{\lambda}_T/2]$. At larger Atwood numbers, a substantial portion of the mixing layer is displaced to the right of $\eta=0$, suggesting that mass conservation is violated when the $x^3$ term is neglected. The apparent displacement of the mole fraction profile can be traced to the symmetry of the normalized diffusivity. Recalling the definition of the dimensionless diffusivity, $\hat{x}^2 = \hat{\varphi}'/\hat{\varphi} = (\ln \varphi)'$, the even symmetry of $\hat{x}$ implies a centered spreading of $\ln \hat{\varphi}$. This is illustrated by the exact collapse and symmetry of the normalized quantity $\ln(R\hat{\varphi})/\ln R$ in Fig.~\ref{fig:profilesNorm_sim}(c), where
\begin{equation}
    \frac{\ln (R\varphi)}{\ln R} = \frac{\ln \bar{\rho}-\ln \rho_L}{\ln \rho_H - \ln \rho_L} 
\end{equation}
is analogous to the mole fraction definition in Eq.~(\ref{eq:molefraction}) but in terms of $\ln \bar{\rho}$ rather than $\bar{\rho}$. The skewness observed in $\hat{\varphi}(\eta)$ therefore arises solely from the exponential mapping of this symmetric, but monotonically increasing function.  

This displacement effect may be quantified using the normalized mass displacement length 
\begin{equation}
    \frac{\hat{\delta}_\eta}{\hat{\lambda}_T}= \frac{R}{R-1} \left[ \int^{1/2}_0 (1-\hat{\varphi}) {\rm d}\left(\frac{\eta}{\hat{\lambda}_T}\right) + \int^0_{-1/2} \left(\frac{1}{R} - \hat{\varphi}\right){\rm d}\left(\frac{\eta}{\hat{\lambda}_T}\right) \right],
    \label{eq:deltanorm}
\end{equation}
where $\hat{\delta}_\eta(R)$ represents the location of a hypothetical sharp interface that would produce the same total mass as the continuous profile $\hat{\varphi}({\eta})$ in an equivalent two-reservoir configuration \citep{goh2025statistically}. Its variation with Atwood number is shown in Fig.~\ref{fig:profilesNorm_sim}(d), which confirms that the relative mass loss increases with Atwood number. We note that a non-zero displacement length arises only in the simplified ODE formulation; the full ODE (\ref{eq:odex2}) is mathematically equivalent to Eq.~(\ref{eq:meandensity3D}) and global mass conservation is easily proven by integrating Eq.~(\ref{eq:meandensity3D}) over the domain.

\begin{figure}
  \begin{minipage}{0.497\textwidth}
    \includegraphics[width=\textwidth]{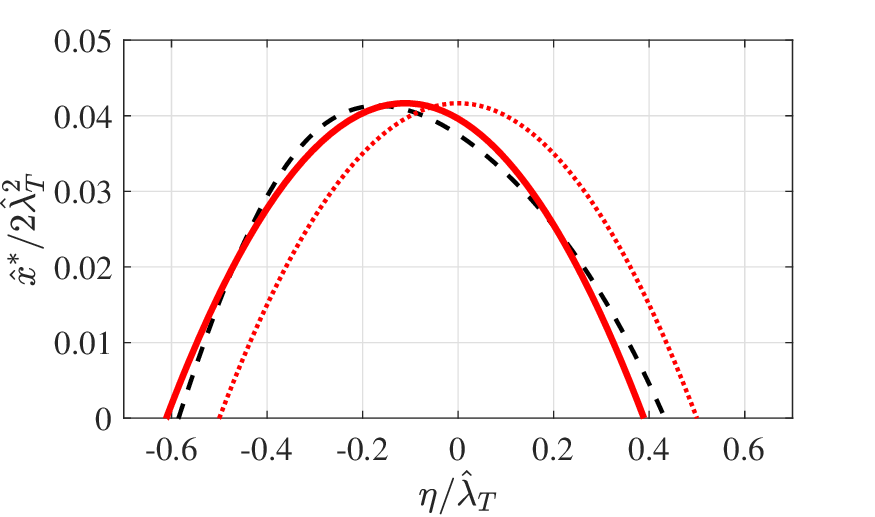}
    {\small (a)}
  \end{minipage}  
  \begin{minipage}{0.497\textwidth}
    \includegraphics[width=\textwidth]{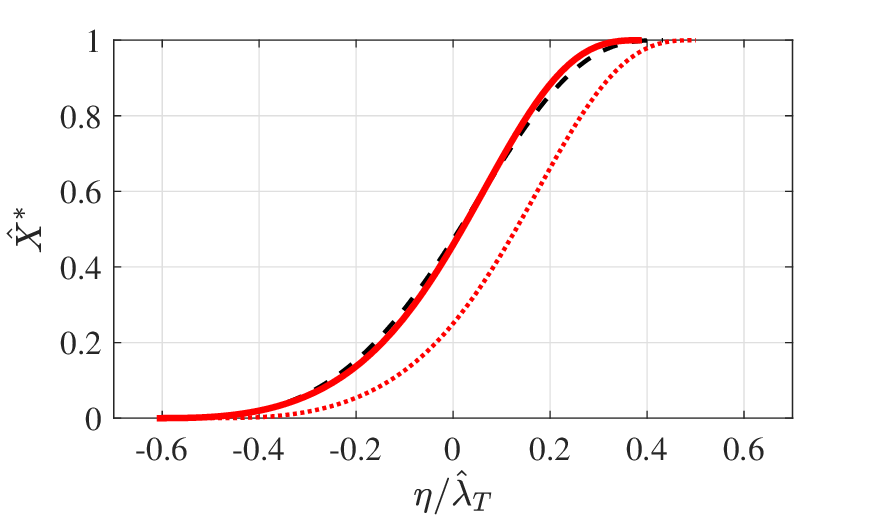}
    {\small (b)}
  \end{minipage}
  \caption{Comparison of the full ODE solution ($x,X$; black dashed), simplified ODE solution ($\hat{x},\hat{X}$; red dotted), and mass-corrected solution ($\hat{x}^*,\hat{X}^*$;  thick red solid) at $A=0.8$: (a) normalized diffusivity/velocity and (b) mole fraction.
  \label{fig:masscorrection}}
\end{figure}

We propose a simple correction to enforce global mass conservation by shifting the profiles leftward by $\hat{\delta}_\eta$. All mass-corrected variables are notated with a ${}^*$ superscript, e.g., $\hat{x}^*(\eta) = \hat{x}(\eta-\hat{\delta}_\eta)$, and the shifted boundaries are $[-\hat{\lambda}_s^*,\hat{\lambda}_b^*]= [-\hat{\lambda}_T/2-\hat{\delta}_\eta,\hat{\lambda}_T/2-\hat{\delta}_\eta]$. Figure \ref{fig:masscorrection} shows the mass-corrected mole fraction for $A=0.8$, compared with the uncorrected solution and the full ODE solution. The mass-corrected profile matches the full ODE solution reasonably well, suggesting that, at least to first order, the inherent profile shape is described well by the simplified ODE, as long as a mass correction is applied after. Representing the turbulent diffusivity as a displaced parabola provides a useful conceptual model for the physics of VD RT mixing. Because $\ln \bar{\rho}$ is the natural variable for describing turbulent mixing in VD flows, diffusion of $\ln \bar{\rho}$ is approximately symmetric and scaled by $\ln R$. A strictly centered spreading of $\ln \bar{\rho}$, however, would violate mass conservation, and must therefore be corrected by a spatial shift toward the light-fluid side. This shift enhances the outward $\ln R$ growth on the spike side while suppressing it on the bubble side, resulting in spikes that grow faster than bubbles.

Motivated by the simplicity and effectiveness of the mass-corrected solution (i.e., $\hat{x}^*(\eta) = \hat{x}(\eta-\hat{\delta}_\eta)$ and $\hat{\varphi}^*(\eta) = \hat{\varphi}(\eta-\hat{\delta}_\eta)$) in approximating the full ODE solution, we seek closed-form expressions for all contributing terms. Although $\hat{x}(\eta)$ and $\hat{\varphi}(\eta)$ are fully specified by Eqs.~(\ref{eq:xsim}) and (\ref{eq:phisim}), the normalized displacement length, defined by Eq.~(\ref{eq:deltanorm}), does not admit an exact closed-form expression and must instead be integrated numerically. We propose an analytical approximation that recovers the expected behavior of the normalized displacement length at both low and high $A$ limits which takes the form
\begin{equation}
    \frac{\hat{\delta}_\eta}{\hat{\lambda}_T} \approx f(A) = \frac{1}{2} \left(\frac{A}{A+c(1-A)^\gamma}\right),
    \label{eq:deltanormfit}
\end{equation}
where $f(0)=0$ and $f(1) = 1/2$ are the expected theoretical limits of $\hat{\delta}_\eta/\hat{\lambda}_T$. The constant $c \approx 4.625$ is fixed by matching the small-$A$ slope, while the exponent $\gamma=0.2932$ is obtained from a least-squares fit over $A \in [0, 1]$. As shown in Fig.~\ref{fig:profilesNorm_sim}(d), the analytical model shows good agreement with the numerical solution across all Atwood numbers. A full set of closed-form expressions for the mass-corrected solution is given by Eqs.~(\ref{eq:xsim}), (\ref{eq:eta0sim}), (\ref{eq:phisim}), and (\ref{eq:deltanormfit}).

\subsection{Mixing layer heights}
\label{sec:heights}
In this section, we consider the scaling of mixing layer heights, first in terms of definitions that encompass the full extent of the mixing layer, followed by the heights of spikes and bubbles.

\subsubsection{Total height}
A key result from \citet{belen1965theory} states that the total mixing layer height scales with $\ln R$. Because this scaling was derived for the solution to the simplified ODE, $\hat{h}_T$, the authors restricted the validity of this result to $R\le 4$ (i.e. $A\le 0.6$). Additionally, it was also assumed in the derivation of Eq.~(\ref{eq:h_smallR}) that the turbulence height $h_t$ scales with $\hat{h}_T$. In this section, we examine the sensitivity of $\ln R$ scaling to these two assumptions, first by employing the numerical solution to the full ODE, and second by considering different height definitions for both the predicted height and the turbulence height. 

\begin{figure}
  \centering
  \begin{minipage}{0.497\textwidth}
    \includegraphics[width=\textwidth]{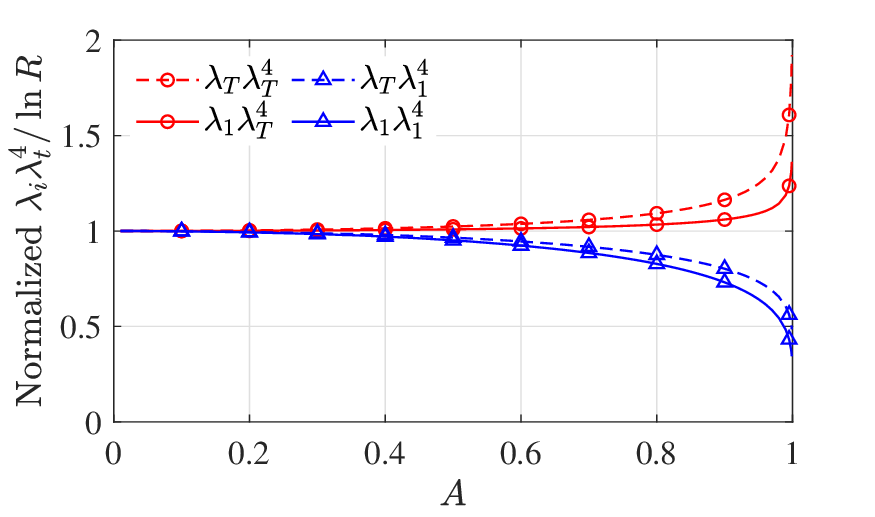}
  \end{minipage}  
  \caption{Comparison of density-ratio scaling $\lambda_i\lambda_t^4$ using different height definitions for $\lambda_i$ and $\lambda_t$. Each scaling is first normalized by $\ln R$ and then by its $A\rightarrow 0$ value. Legend: predicted height $\lambda_i = \lambda_T$ (dashed line), $\lambda_1$ (solid line); turbulence height $\lambda_t = \lambda_T$ (red circles), $\lambda_1$ (blue triangles)
  \label{fig:modelvslnR}}
\end{figure}

The general scaling relationship for any height is Eq.~(\ref{eq:h_general}), with density-ratio effects captured via $\lambda_i\lambda_t^4$. This factor represents the scaling of the predicted height $h_i$ assuming that the turbulence height governing the underlying dynamics of the mixing layer is $h_t$. Using the full ODE solution, the density-ratio scaling factor is computed by considering two different height definitions, $\lambda_1 = \int 4 X(1-X){\rm d}\eta$ and $\lambda_T = \lambda_s+\lambda_b$, as possible candidates for both the predicted height $\lambda_i$ and turbulence height $\lambda_t$. Each version of $\lambda_i\lambda_t^4$ is first divided by $\ln R$, then normalized by its $A\rightarrow 0$ value, and shown in Fig.~\ref{fig:modelvslnR} as a function of Atwood number.

For a fixed choice of turbulent height $\lambda_t$, the scaling of $h_T\sim\lambda_T\lambda_t^4$ increases more rapidly with Atwood number than $h_1\sim\lambda_1\lambda_t^4$, indicating that localized growth at the edges (as opposed to a spatially averaged growth) becomes increasingly significant at high $A$. For a given $h_i$, the sensitivity to the choice of $\lambda_t$ is amplified by its quartic dependence, where $\lambda_i\lambda_1^4$ scaling is sub-logarithmic in $R$, while $\lambda_i\lambda_T^4$ scaling is super-logarithmic. All cases shown in Fig.~\ref{fig:modelvslnR} remain within 20\% of $\ln R$ scaling up to $A\approx 0.84$ (or $R\approx 11$), but diverge significantly as $A\rightarrow 1$. The physically appropriate choice of $h_t$ remains an open question. Given the approximations in the theoretical model, the divergent behavior of $\lambda_i\lambda_1^4$ and $\lambda_i\lambda_T^4$ scalings, and the lack of converged high-Atwood-number data for validation, we propose that $\ln R$ scaling may be used as a ``median'' estimate across all Atwood numbers, at least until further developments become available. As shown in \citet{goh2026self}, $\ln R$ scaling lies well within the observed scatter from available numerical and experimental data, supporting its applicability over a wide range of Atwood numbers.

\subsubsection{Spikes and bubbles}
It is common in the VD RT literature to track the growth of spikes and bubbles separately. As discussed in the introduction, it has been generally observed that bubble heights scale linearly with Atwood number while spike heights do not; this translates to near-constant values of $\alpha_b(A)$ but growing $\alpha_s(A)$ when the Boussinesq-derived scaling of Eq.~(\ref{eq:Agt2}) is applied \citep{Dimonte2000,youngs2013density,banerjee2010detailed,zhou2019time}. We seek to interpret these empirical observations using the theoretical model, by comparing $\alpha_{s/b}$ values from both the full and mass-corrected models with available RT data.

\begin{figure}
  \begin{minipage}{0.497\textwidth}
    \includegraphics[width=\textwidth]{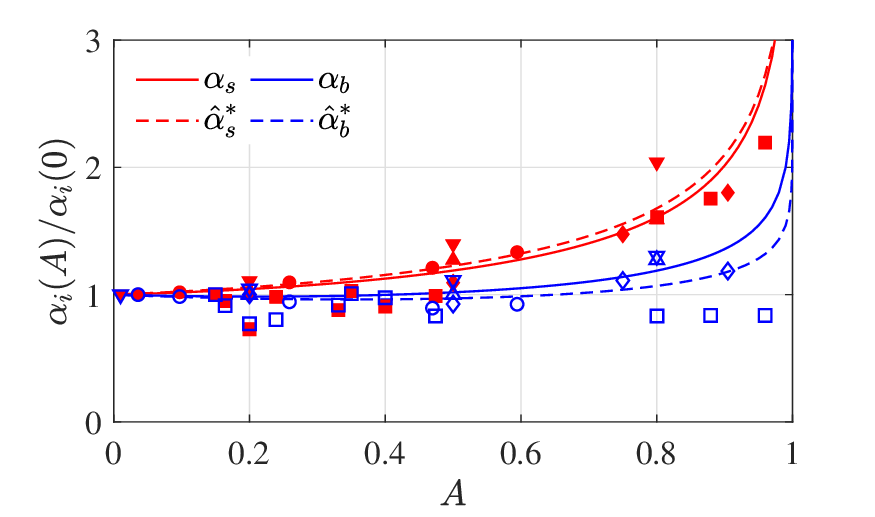}
    {\small (a)}
  \end{minipage}
  \begin{minipage}{0.497\textwidth}
    \includegraphics[width=\textwidth]{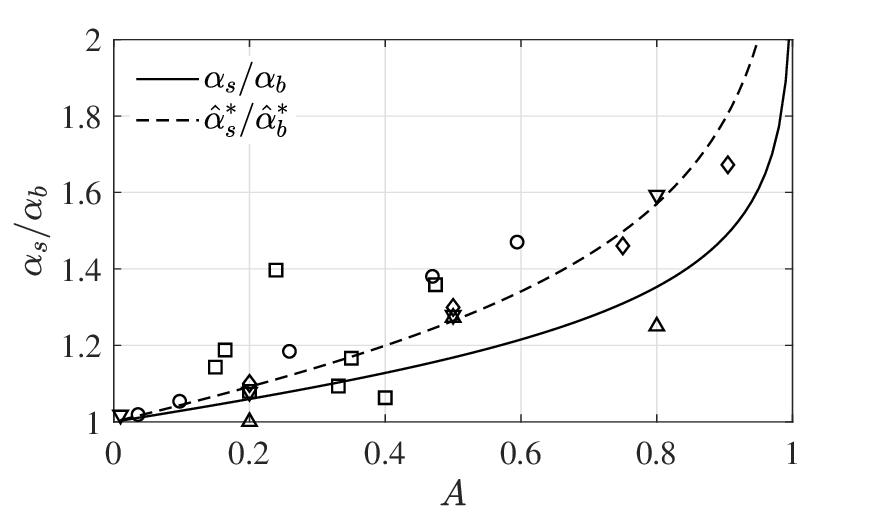}
    {\small (b)}
  \end{minipage}
  \caption{(a) Normalized growth parameters for spike (red) and bubble (blue) heights; and (b) ratio of spike to bubble heights. Legend: full solution $\alpha_i=\alpha_{s/b}$ (solid line) and mass-corrected solution $\alpha_i=\hat{\alpha}^*_{s/b}$ (dashed line); symbols represent equivalent RT data from \citet{Dimonte2000} ($\square$), \citet{banerjee2010detailed} ({\large $\circ$}), \citet{youngs2013density} ($\lozenge$), \citet{zhou2019time} ($\triangle$), and \citet{goh2026self} ($\triangledown$).
  \label{fig:bubblespike}}
\end{figure}

Growth parameters $\alpha_i$ are derived from the theoretical model by equating Eqs.~(\ref{eq:Agt2}) and (\ref{eq:h_general}) to give
\begin{equation}
    \alpha_i = \frac{K^2\lambda_i\lambda_t^4}{A} = \frac{K^2\lambda_i(270 \ln R)^{4/5}}{A},
\end{equation}
where a fixed turbulence height, $\lambda_t=\hat{\lambda}_T$, is used throughout this section.
For ease of comparison, all $\alpha_i$ values are normalized by their $A\rightarrow 0$ value; a normalized value of 1 corresponds to exact $Agt^2$ scaling. In Fig.~\ref{fig:bubblespike}(a), we compare two spike-bubble pairs corresponding to $\alpha_{s/b}$ and $\hat{\alpha}^*_{s/b}$. For both the full ODE solution and the mass-corrected solution, bubble heights show better agreement with $Agt^2$ scaling than spike heights, although bubble heights do also eventually deviate significantly from linear Atwood-number scaling beyond $A \approx 0.8$. This difference in scaling between spikes and bubbles results in a growing $\alpha_s/\alpha_b$ ratio, as shown in Fig.~\ref{fig:bubblespike}(b). Equivalent RT data from experiments and simulations are shown in both Figs. \ref{fig:bubblespike}(a,b). Due to the presence of diffuse interfaces and statistical noise, there is some ambiguity in identifying spike/bubble fronts and each study uses a different method or threshold. Nonetheless, the turbulent diffusivity model clearly captures the trend and makes quantitative predictions that are within the spread of values observed from physical RT studies. 

\section{Summary and discussion}
\label{sec:discussion}

Despite its relevance and simplicity, the proposed turbulent diffusivity model from \citet{belen1965theory}, along with the associated $\ln R$ scaling, has seen limited adoption in the RT literature. In the present work, we have streamlined the original derivation, validated the model, extended the analysis, and contextualized the results within more recent developments in RT turbulence.

The full self-similar ODE captures many qualitative features of non-Boussinesq RT mixing layers, namely
\begin{enumerate}
    \item The total mixing layer height scales approximately with $\ln R $, rather than $A$.
    \item Spike heights appear to deviate more strongly from $Agt^2$ scaling than bubble heights.
    \item The ratio of spike to bubble heights increases with Atwood number.
    \item Velocity profiles (represented by $x$) are shifted toward the light-fluid side.
\end{enumerate}
These behaviors are inherently captured by the model of \citet{belen1965theory}; however, with the exception of observation (1), they were not discussed in the original paper. The present work brings these results to light and highlights that many now-known features of non-Boussinesq RT flows are already contained within a simple 1D theoretical framework that has long been available but whose relevance has perhaps been overlooked.

Although the full ODE (\ref{eq:odex2}) is straightforward to solve numerically, we propose a mass correction to the simplified ODE solution and demonstrate that it reproduces the same qualitative characteristics. In this formulation, the normalized turbulent diffusivity is represented exactly by a downward parabola following $\ln R$ scaling, with a spatial shift imposed by mass conservation. This further simplification offers two main advantages. First, the simple representation of the turbulent diffusivity as a spatially-shifted parabola may facilitate analytical developments of Reynolds-averaged turbulence models (e.g., \citet{schilling2021self}); a full set of closed-form expressions for this representation is provided in Sec.~\ref{sec:profiles_sim}. Second, the model highlights the competing dynamics between diffusion of $\ln \bar{\rho}$ and mass conservation, offering a mechanistic explanation for the observed asymmetry in non-Boussinesq RT mixing layers.


\begin{acknowledgments}
This material is based upon work supported by the National Science Foundation under Grant No. 2422513.
This work used resources of the National Energy Research Scientific Computing Center, a DOE Office of Science User Facility supported by the Office of Science of the U.S. Department of Energy under Contract No. DE-AC02-05CH11231 using NERSC award FES-ERCAP0031321. The authors also acknowledge the use of Google AI to assist in translating the original document of \citet{belen1965theory} from Russian into English.
\end{acknowledgments}

\bibliography{references}

\end{document}